\documentclass[letterpaper,11pt]{emulateapj}
\usepackage{epsfig}
\usepackage{natbib}
\usepackage{lscape}
\begin{document}
\newcommand{\kms}{km~s$^{-1}$}
\newcommand{\Msun}{M_{\odot}}
\newcommand{\Lsun}{L_{\odot}}
\newcommand{\ML}{M_{\odot}/L_{\odot}}
\newcommand{\etal}{{et al.}\ }
\newcommand{\hhh}{h_{100}}
\newcommand{\hsq}{h_{100}^{-2}}
\newcommand{\tn}{\tablenotemark}
\newcommand{\mdot}{\dot{M}}
\newcommand{\p}{^\prime}
\newcommand{\kmsMpc}{km~s$^{-1}$~Mpc$^{-1}$}

\title{The Extragalactic Distance Database: Color--Magnitude Diagrams}

\author{Bradley A. Jacobs}
\affil{Institute for Astronomy, University of Hawaii, 2680 Woodlawn Drive, Honolulu, HI 96822}

\author{Luca Rizzi}
\affil{United Kingdom Infrared Telescope, 660 N. A'ohoku Pl, Hilo, HI 96720}

\author{R. Brent Tully,}
\affil{Institute for Astronomy, University of Hawaii, 2680 Woodlawn Drive, Honolulu, HI 96822}

\author{Edward J. Shaya}
\affil{Department of Astronomy, University of Maryland, College Park, MD 20742}

\author{Dmitry I. Makarov}
\affil{Isaac Newton Institute of Chile, SAO Branch, Russia \\
Special Astrophysical Observatory, Nizhniy Arkhyz, Karachai-Cherkessia 369167, Russia}

\and

\author{Lidia  Makarova}
\affil{Isaac Newton Institute of Chile, SAO Branch, Russia \\
Special Astrophysical Observatory, Nizhniy Arkhyz, Karachai-Cherkessia 369167, Russia}

\begin{abstract}
The CMDs/TRGB (Color-Magnitude Diagrams/Tip of the Red Giant Branch) section of the Extragalactic Distance Database contains a compilation of observations of nearby galaxies from the Hubble Space Telescope.   Approximately 250 (and increasing) galaxies in the Local Volume have CMDs and the stellar photometry tables used to produce them available through the web.  Various stellar populations that make up a galaxy are visible in the CMDs, but our primary purpose for collecting and analyzing these galaxy images is to measure the TRGB in each.  We can estimate the distance to a galaxy by using stars at the TRGB as standard candles.  In this paper we describe the process of constructing the CMDs and make the results available to the public.
\end{abstract}


\section{Introduction}
One of the most productive means of measuring distances to nearby galaxies over the last 10-15 years has been to use stars at the Tip of the Red Giant Branch (TRGB) as standard candles.  As the name implies, this technique requires accurate photometry of RGB stars in other galaxies, and the facility most suited to this purpose is the Hubble Space Telescope (HST).  We have compiled HST observations of $\sim$250 (and counting) nearby galaxies.  These observations include imaging in two colors for each galaxy, and we present the resulting Color-Magnitude Diagrams (CMD) as part of the Extragalactic Distance Database\footnote{Located at: http://edd.ifa.hawaii.edu/ under the Miscellaneous Distances:CMDs/TRGB tab.} \citep{tul09}.
\par
The CMDs/TRGB index page of the database lists each galaxy's PGC/LEDA number (Principal Galaxy Catalog/Lyon Meudon Extragalactic Database), common name, HST observing program number, and camera used.  Each entry is indexed by its PGC/LEDA number rather than its common name in order to prevent a single galaxy from being listed twice in the database.  The PGC/LEDA number and common name fields both serve as links to more detailed information about the galaxy.  Each galaxy's CMD webpage is accessible by clicking its common name, while clicking the PGC/LEDA number sends one to a DSS2 image of the galaxy provided by HyperLEDA\footnote{Located at: http://leda.univ-lyon1.fr/} \citep{pat03}.
\par
Though our primary purpose for producing these CMDs is to measure the apparent magnitude of each galaxy's TRGB, there is a great deal of information about stellar populations that can be gleaned from the figures and photometry tables from which they are derived.  For example, a galaxy's metal abundance can be estimated by the color of the  RGB \citep{dac90, lee93}.  Some of the HST observing programs from which we present our reductions were designed specifically for the purpose of measuring the TRGB in many galaxies, others came from case studies of a single galaxy, and we include observations from these and various other programs that happened to include observations in filters appropriate for locating the TRGB.  The CMD database consists of as large a sample as available of galaxies within $\sim$10 Mpc, each analyzed in a consistent manner.

\section{Data}
The CMDs are produced by performing photometry on HST images taken with either the Wide Field-Planetary Camera 2 (WFPC2) or the Advanced Camera for Surveys (ACS).  The HST flight filter most appropriate for locating the TRGB is $F814W$, which can be described as `wide $I$.'  In order to discriminate between the RGB and other star types we require observations in another filter.  Either of the $V$-equivalent filters: $F606W$ and $F555W$ most commonly serve this purpose, but it is possible to use one of the $B$-like filters such as $F475W$ as well.  All of the images used to produce CMDs are available to the public through the HST Archive.\footnote{Located at: http://archive.stsci.edu/hst/}

\section{CMDs/TRGB Catalog}
In addition to the general identification, linking, and observational information provided for each galaxy in the CMDs/TRGB Catalog, a subset of the galaxy entries contain data related to and including a TRGB distance measurement.  The eventual goal is to publish this information for each galaxy in the catalog.  An extraction of the catalog with currently available data for 97 galaxies is given in Table \ref{db}.  The bulk of the columns in the catalog contain TRGB magnitude and color information.  The first three columns of data list our best estimate of the $F814W$ magnitude of the TRGB, followed by low and high values representing the bounds of the 68\% confidence interval.  These magnitudes are followed by a measurement of the color of the TRGB in the HST flight filters, again with a best, low, and high value reported.  In order to facilitate comparison to other datasets we also convert these values to Johnson-Cousins $V$ and $I$, using the prescriptions from \citet{dola} and \citet{hol95} for WFPC2, and \citet{sir05} for ACS.   Next, we present an estimate of the foreground extinction based on the dust maps of \citet{sch98}, and use this in our calculations of each galaxy's distance modulus, which we list after the extinction values along with their confidence bounds. 

\section{Footprint Images}
\begin{figure}[!ht] 

\begin{center} 

\includegraphics[scale=.5]{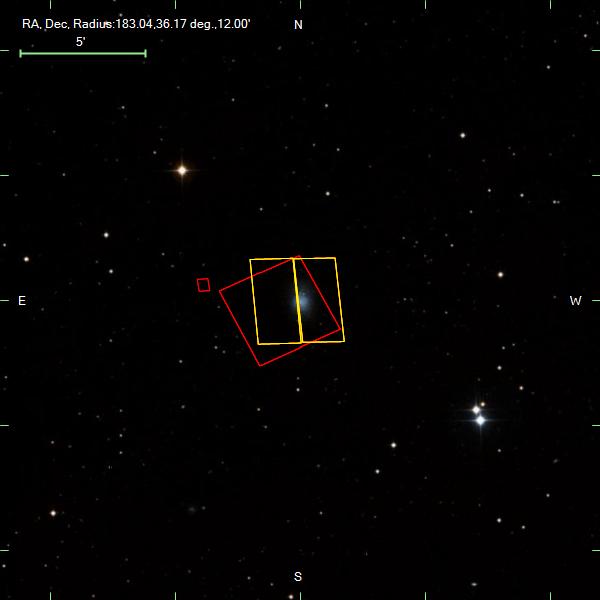}

\vspace{.1in}
\includegraphics[scale=.5]{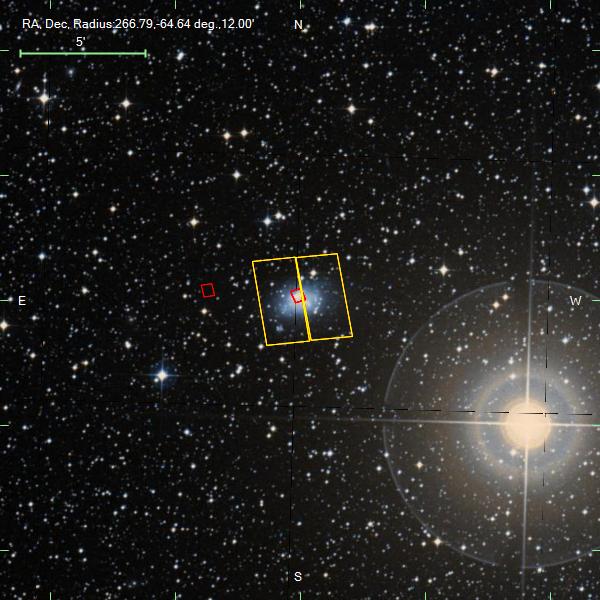}
\caption{HST Footprints of the galaxies NGC4163 (upper) and IC4662 (lower) from ACS observations.  The footprint of observations from program 9771 is shown in yellow.  The red boxes denote the footprint from another ACS observation of the field.}
\label{foot}
\end{center}
\end{figure}

The individual galaxy webpages linked from the catalog page each display the footprint of the HST observations used in our analysis.  These images were obtained through the Hubble Legacy Archive.\footnote{Located at: http://hla.stsci.edu/}  We present them in order to put the observations into context.  Where there are multiple observations of galaxies of large angular extent  we prefer those located towards the edge of the galaxy as seen in the footprint image (taken from Digitized Sky Survey-2), since it is such regions where low-metallicity RGB stars most suitable for TRGB measurements are common and crowding is less of a problem than towards the core.  Many galaxies in the sample have a small angular size that fits well within the field-of-view of ACS or WFPC2, and as such no attempt can be made to focus on the halo stars.  See for example NGC4163 and IC4662 in Figure \ref{foot}.  The footprints of the observations used in our analysis are highlighted in yellow.  The red boxes represent the footprints of other ACS observations in the field.  Footprints of observations other than those used for the production of the galaxy's CMD often appear in these images, but those that are used are always highlighted in yellow.  In order to reduce the clutter caused by the necessity of showing these unused observations we only show the footprints of one instrument on each image.  So each CMD used with ACS data has one corresponding yellow ACS footprint, and any other ACS observations that happen to fall on the footprint image are shown in red, but no WFPC2 footprints are shown.  The situation is similar for galaxies observed with WFPC2, except that non-used WFPC2 observations are shown in blue, and the ACS footprints are turned off.  Each image in the database is set to a consistent zoom level corresponding to a field of view of $24'\times 24'$.

\section{Observation Images}
\begin{figure}[!ht] 

\begin{center} 

\includegraphics[scale=.75]{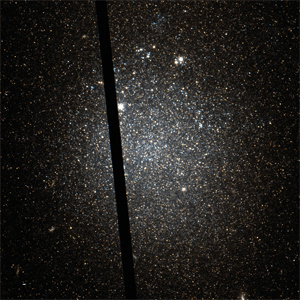}\\

\vspace{.1in}
\includegraphics[scale=.75]{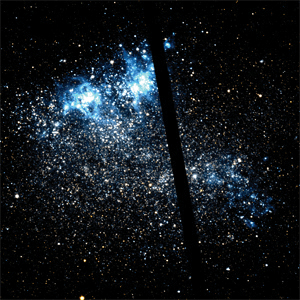}
\caption{Pseudo-color images of NGC4163 (upper) and IC4662 (lower) produced from $F606W$ and $F814W$ filters.}
\label{col}
\end{center}
\end{figure}

For a subset of galaxies in the database we combine the two-filter observations into psuedo-color images in order provide a sense of the galaxies appearance (we plan to eventually post these images for every galaxy in the catalog).  We denote these images at pseudo-color because we lack three channel color information and map the shorter wavelength filter (typically $F606W$) to the blue channel and the $F814W$ image to the red channel.  For the green channel we take an average of the two filters.  The drizzled ACS images are suitable for this process, while WFPC2 images require additional processing.  We use the `hstmosaic' routine that is part of the HSTPHOT photometry package from \citet{dol00} to align WFPC2's chips, and clean the image.   Figure \ref{col} shows the results for ACS observations of NGC4163 and IC4662.  These galaxies were chosen in part for their differences in stellar populations even though both have a prominent RGB, which will be evident in their CMDs shown below.  There are regions suggestive of active star formation in the image of IC4662 that are not apparent in NGC4163.  Inspection of the images allows one to identify regions dominated by old, intermediate, and ongoing star formation.  There is no rigor to the detailed color mapping.

\section{Analysis}
We perform photometry on the WFPC2 images using the HSTPHOT \citep{dol00} software package, which was developed specifically for observations of resolved stellar populations using this instrument.  Dolphin expanded HSTPHOT into a more general photometry software program called DOLPHOT.\footnote{Available at: http://purcell.as.arizona.edu/dolphot/}  This release includes an ACS module, and we use this package for analysis of images taken with ACS.   Each CMD webpage in the database contains a link to the photometry file output from HSTPHOT or DOLPHOT.  These tables hold information about the measurement of a single star per row.  There are, among others, columns that describe a star's position on the image, and its apparent magnitude in both flight and ground-based filters, as well as several characterizations of the quality of the measurement.  If there are several images per filter available then these values are displayed for each individual image as well as in combination.  Hence it is not uncommon for the photometry table to contain more than fifty columns.  Though the output is quite complex it breaks into simple sections that are described in detail in the HSTPHOT and DOLPHOT manuals.  The column definitions vary between HSTPHOT (for WFPC2) and DOLPHOT (for ACS), so it is  important to refer to the correct manual when interpreting the photometry files.  To facilitate this we list which camera was used in the observations on the CMDs/TRGB catalog page.
\par
The full photometry files are available through the CMD webpage, but we do not plot every star in those files on their respective CMDs.  We use a set of the data quality parameters to reject some measurements as unreliable.  Each measurement has a $\chi$ value, Sharpness parameter, Object Type, and PSF parameter assigned to it.  In order for the measurement to be accepted we require that it fall in the range of values recommended in the manuals, as well as have a Signal-to-Noise ratio: $S/N \geq 5.$  Each star measurement that passes all these tests is plotted on the CMD shown on each galaxy's webpage.

\section{Color-Magnitude Diagrams}
\begin{figure}[!ht] 

\begin{center} 

\includegraphics[scale=0.46]{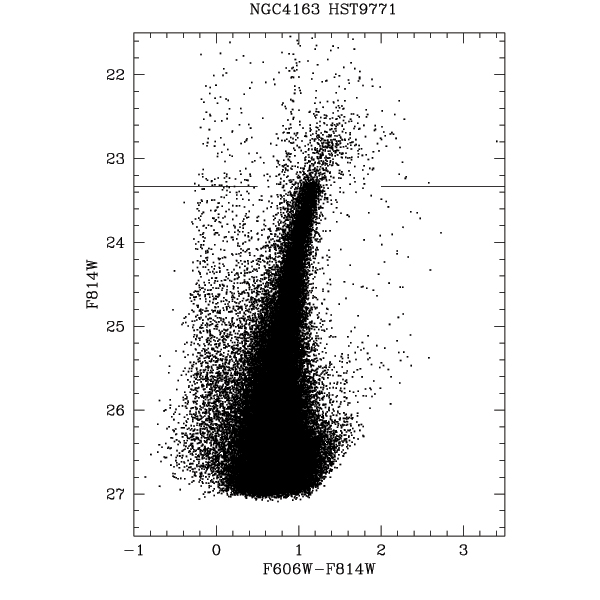}
\includegraphics[scale=0.46]{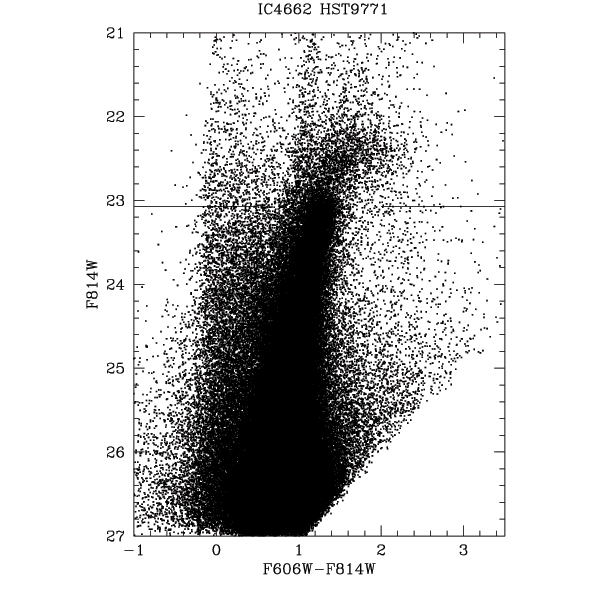}
\caption{CMDs of the galaxies NGC4163 and IC4662 in $F606W - F814W$ versus $F814W$ constructed from ACS observations with HST observing program: 9771.  The broken horizontal lines at $F814W=23.33$ and $F814W=23.07$ in NGC4163 and IC4662's respective CMDs mark the magnitude of their TRGB.}
\label{n4163}
\end{center}
\end{figure}

Figure \ref{n4163} shows example CMDs from ACS observations of NGC4163 and IC4662 in the $F814W$ and $F606W$ filters.  The most prominent feature on the CMDs of these galaxies (and of most other galaxies in the database) is the RGB, seen here as solid wedges due to the high density of points.   Other populations visible include the Main Sequence and Blue Loop stars that together form a vertical band seen between about $-0.4 < F606W-F814W < +0.4$.  Above and redward of the TRGB are the Asymptotic Giant Branch stars, which are centered at about $F814W=23$ and $F606W-F814W=+1.3$ for NGC4163 and $F814W=22.6$ and $F606W-F814W=+1.5$ for IC4662.  Each of these regions of the CMD hold a wealth of information about the stellar content and history of the galaxies, but here we focus our attention on the RGB and its tip, which we denote with broken horizontal lines at $F814W=23.33$ for NGC4163 and $F814W=23.07$ for IC4662.  Note that although the stars that make up NGC4163 and IC4662 are clearly different in metallicity and likely age as well, the TRGB is prominent in both.

\section{TRGB Distances}
Red Giant stars burn hydrogen in shells around a degenerate helium core that has a temperature independent equation of state.  While a star is in this phase the mass and temperature of the core continuously increase until the temperature reaches about $10^8$ K, at which point triple-$\alpha$ helium burning ignites and sends the star off the RGB to the Horizontal Branch.  The helium falling on the core from the outer shell drives this increase in temperature, and hence the degeneracy is broken when the core mass crosses the threshold for helium burning.  This mass threshold, therefore, determines an upper limit to the luminosity of Red Giants.  The flux from these stars is least sensitive to age and metallicity in $I$-band \citep{riz07}, thus our requirement that galaxies be observed in $F814W$.
\par
\begin{figure}[!ht] 

\begin{center} 

\includegraphics[scale=0.6]{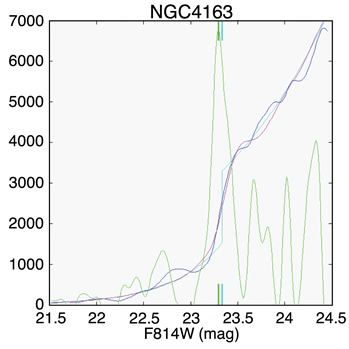}
\includegraphics[scale=0.6]{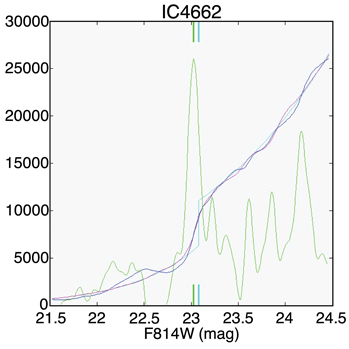}
\caption{Histogram (blue) and fits to the CMDs of NGC4163 and IC4662 taken from a screenshot of TRGBTOOL's interface.  The magenta curve is the model luminosity function convolved with photometric errors and incompleteness, from this the intrinsic luminosity function is computed and plotted in cyan.  The green curve represents the first derivative of the histogram of the stellar luminosity function.  The cyan tick on the $F814W$ axis marks the magnitude of the TRGB, and the green tick marks peak of the first derivative of the histogram.  For NGC4163 the peak of the derivative and the TRGB are at$F814W=23.30$ and $F814W=23.33$, respectively, and are at $F814W=23.02$ and $F814W=23.07$ for IC4662.}
\label{ttool}
\end{center}
\end{figure}
\begin{figure}[!ht] 

\begin{center}

\includegraphics[scale=0.45]{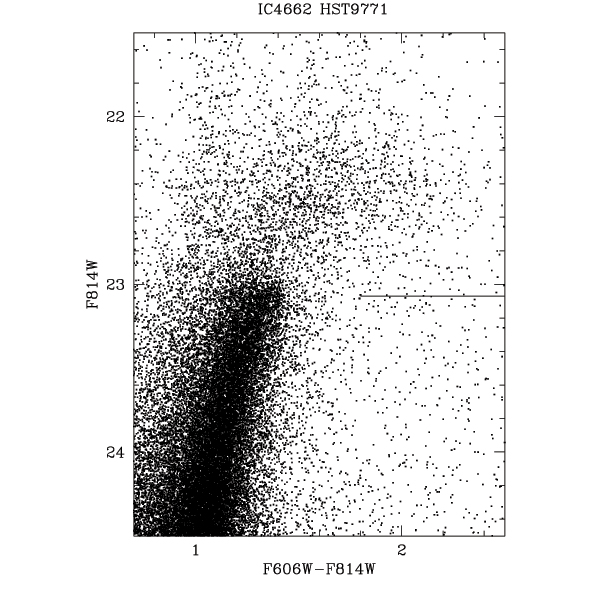}
\caption{CMD of IC4662 in $F606W - F814W$ versus $F814W$ showing the same data as in Figure \ref{n4163} zoomed to a smaller area of color-magnitude space.  The line segment to the right of the RGB denotes the tip measurement of $F814W=23.07$.}
\label{zoom}
\end{center}
\end{figure}

Early attempts to use the TRGB as a distance indicator simply used by-eye estimates to measure its apparent magnitude.  Over the years TRGB measurement methods have become more sophisticated and reproducible \citep[e.g.][]{lee93, men02, mak06, mag08}.  We employ a program called TRGBTOOL that uses a maximum-likelihood method described by \citet{mak06}.  This program fits a power-law with a step to an unbinned histogram of the number of stars versus magnitude.  This is done because, as seen in Figure \ref{n4163} (and indicated by a broken horizontal line), the TRGB appears as a discontinuity in the density of stars at a particular magnitude, decreasing from the tip location to slightly brighter magnitudes.  This method also incorporates the results of artificial star tests performed on the HST images using HSTPHOT or DOLPHOT.  This fake photometry is used to account for how completeness, crowding, and color-spreading affect the reliability of the tip measurement. Figure \ref{ttool} shows the histograms (blue) and fits (cyan and magenta) to the CMDs of NGC4163 and IC4662 taken from a screenshot of TRGBTOOL's interface.  The green curve shows the first derivative of the histogram of the stellar luminosity function.  The magenta curve is the model luminosity function convolved with photometric errors and incompleteness.  We take the location of the discontinuity in the cyan intrinsic luminosity function as the magnitude of the TRGB.  For the galaxy NGC4163 the case for the TRGB to be located at $F814W=23.33\pm0.02$ is clear.  The best fit power-law discontinuity falls very close to the steepest part of the rise in the stellar luminosity function.  This can be seen in Figure \ref{ttool} where the peak of the derivative of the stellar luminosity function is marked by a green tick and the TRGB discontinuity is marked by a cyan tick.  IC4662's fit at $F814W=23.07\pm0.02$ is not so clear in Figure \ref{n4163}.  The CMD is densely populated around the tip.  This situation is clarified by plotting a zoomed-in version of the CMD, shown in Figure \ref{zoom}.

\par
Once we have determined the apparent magnitude of a particular galaxy's TRGB we are then able to calculate a distance modulus.  The absolute magnitude of stars at the TRGB is $M_I \approx -4.0$ with only a weak dependence on age and metallicity in this wavelength range.  \citet{riz07} present a zero-point calibration that is adjusted for metallicity as well as for the details of HST observations.  We use their calibration of the absolute magnitude and correct for foreground extinction using values derived from the dust maps by \citet{sch98} to produce a distance modulus using the following equation: 
\begin{displaymath}
DM=m_{TRGB}-A_I+4.05-0.217[(V-I)-(A_V-A_I)-1.6] 
\end{displaymath}
where $m_{TRGB}$ is the apparent magnitude of the TRGB in $I$-band, $(V-I)$ is the color of the TRGB, and $A_V$ and $A_I$ are the respective foreground extinction terms.   \citet{riz07} present analogous formulae for calculating the distance modulus using HST flight filter magnitudes.  Extinction within the target galaxies cannot be estimated with these observations. However, we note that the stars most amenable to TRGB distance measurements are old and low-metallicity, those that are preferentially found in the outer halo where dust obscuration is minimal, so when possible we focus on these regions.  The uncertainty of TRGB distances we measure from galaxies with well-sampled CMDs (those with a tip at least 1.5 mag above the limit of the observations) typically hover around a three percent.  For comparison the interval between tick marks in the vertical scale of the CMDs corresponds to 10\% shifts in distance.  Details of our procedures are described by \citet{mak06} and \citet{riz07}.  For galaxies with a TRGB measurement on the CMDs/TRGB webpage there is also a distance modulus.  A portion of this webpage is reproduced in Table \ref{db}.
   
\section{Acknowledgments}
Support for this project has come from HST programs: AR-9950, GO-9771, GO-10210, GO-10235, GO-10905, and AR-11285.  DIM and LM acknowledge support from Russian Foundation for Basic Research grant 08-02-00627.  Some of the data presented in this paper were obtained from the Multimission Archive at the Space Telescope Science Institute (MAST). STScI is operated by the Association of Universities for Research in Astronomy, Inc., under NASA contract NAS5-26555. Support for MAST for non-HST data is provided by the NASA Office of Space Science via grant NAG5-7584 and by other grants and contracts.  Footprint images were obtained from the Hubble Legacy Archive, which is a collaboration between the Space Telescope Science Institute (STScI/NASA), the Space Telescope European Coordinating Facility (ST-ECF/ESA) and the Canadian Astronomy Data Centre (CADC/NRC/CSA).

\section{Table}
Table \ref{db} (beginning on the following page) lists the results for galaxies for which we have measured a tip and calculated a distance modulus.  The first seven columns list: PGC Number, Common Name, HST observing program number, camera used, and TRGB magnitude in $F814W$, followed by its 68\% lower and upper limits.  This pattern of best mag, lower limit mag, and upper limit mag continues for the rest of the table, with the exception of the reddening column E($B$--$V$), and are presented in blocks of three.  Following the $F814W$ TRGB magnitude block are the $F606W$--$F814W$ colors, $F555W$--$F814W$ colors, the $I$-band TRGB magnitudes, and the $V$--$I$ colors.  These are followed by reddening magnitude E($B$--$V$) column, and a block of distance moduli.

\clearpage

\LongTables
	\begin{landscape}
\begin{deluxetable}{rlcc|ccc|ccc|ccc|ccc|ccc|c|ccc} %NOTE the double line '||' only makes a line appear in my version of Latex.
\tabletypesize{\scriptsize}

 \tablecaption{CMD/TRGB}
 \tablewidth{0pt}
\tablehead{
\colhead{PGC}&\colhead{Name}&\colhead{GO\#}&\colhead{Camera}&\colhead{T814}&\colhead{Lo}&\colhead{Hi}&\colhead{6-8}&\colhead{Lo}&\colhead{Hi}&\colhead{5-8}&\colhead{Lo}&\colhead{Hi}&\colhead{Tip I}&\colhead{Lo}&\colhead{Hi}&\colhead{V-I}&\colhead{Lo}&\colhead{Hi}&\colhead{E(B-V)}&\colhead{D M}&\colhead{Lo}&\colhead{Hi}
}
\startdata
     143&DDO221& 6813&  WFPC2&20.99&20.90&21.02&     &     &     & 1.49& 1.40& 1.52&20.95&20.86&20.98& 1.51& 1.42& 1.54&0.036&24.95&24.86&25.00\\
     621&ESO349-031& 9771&    ACS&23.47&23.43&23.50& 1.08& 1.08& 1.10&     &     &     &23.47&23.44&23.51& 1.42& 1.41& 1.43&0.012&27.53&27.49&27.57\\
     930&   NGC45& 9774&    ACS&25.04&24.96&25.11&     &     &     & 1.49& 1.47& 1.52&24.94&24.92&24.97& 1.58& 1.56& 1.60&0.020&29.10&29.02&29.18\\
    1014&NGC55& 8697&  WFPC2&22.66&22.63&22.68&     &     &     & 1.68& 1.66& 1.70&22.62&22.59&22.65& 1.70& 1.69& 1.72&0.013&26.62&26.59&26.65\\
    1038&    ESO410-005&10503&    ACS&22.37&22.33&22.40& 1.13& 1.12& 1.14&     &     &     &24.52&24.47&24.56& 1.48& 1.46& 1.49&0.014&26.42&26.38&26.45\\
    1305&IC10& 6406&  WFPC2&21.93&21.88&22.04&     &     &     & 2.80& 2.73& 2.73&21.95&21.90&22.06& 2.84& 2.77& 2.77&1.284&23.52&23.48&23.64\\
    1641&    ESO294-010&10503&    ACS&22.38&22.35&22.43& 1.16& 1.13& 1.17&     &     &     &23.81&23.78&23.84& 1.51& 1.47& 1.53&0.006&26.45&26.41&26.50\\
    2004&NGC147& 6233&  WFPC2&20.69&20.64&20.83&     &     &     & 2.05& 1.98& 2.20&20.67&20.63&20.81& 2.08& 2.00& 2.23&0.173&24.32&24.25&24.47\\
    2329&NGC185& 6699&  WFPC2&20.42&20.40&20.44&     &     &     & 2.00& 1.97& 2.03&20.40&20.38&20.42& 2.03& 2.00& 2.06&0.181&24.04&24.02&24.07\\
    2429&NGC205& 6699&  WFPC2&20.70&20.68&20.71&     &     &     & 2.01& 1.99& 2.03&20.67&20.66&20.69& 2.04& 2.01& 2.06&0.079&24.49&24.47&24.51\\
    2578&DDO226& 8192&  WFPC2&24.44&24.32&24.55& 1.01& 0.98& 1.03&     &     &     &24.40&24.28&24.52& 1.44& 1.41& 1.48&0.016&28.45&28.33&28.58\\
    2758&  NGC247&10915&    ACS&23.88&23.85&23.90& 1.39& 1.38& 1.40&     &     &     &25.06&24.93&25.19& 1.79& 1.78& 1.81&0.018&27.88&27.85&27.90\\
    2789&  NGC253&10523&    ACS&23.84&23.82&23.86& 1.49& 1.46& 1.51&     &     &     &25.13&25.01&25.24& 1.91& 1.87& 1.93&0.020&27.82&27.79&27.84\\
    2881&    ESO540-030&10503&    ACS&23.71&23.67&23.74& 1.12& 1.10& 1.13&     &     &     &24.44&24.40&24.48& 1.46& 1.43& 1.47&0.023&27.75&27.71&27.79\\
    2902&DDO6& 8192&  WFPC2&23.63&23.55&23.70& 0.96& 0.91& 1.00&     &     &     &23.59&23.51&23.66& 1.38& 1.30& 1.43&0.017&27.67&27.57&27.75\\
    2933&    ESO540-032&10503&    ACS&23.75&23.72&23.78& 1.14& 1.13& 1.15&     &     &     &23.96&23.95&23.98& 1.48& 1.47& 1.50&0.020&27.79&27.76&27.82\\
    3238&  NGC300&10915&    ACS&22.59&22.55&22.64& 1.44& 1.40& 1.46&     &     &     &22.61&22.57&22.66& 1.85& 1.81& 1.87&0.013&26.59&26.54&26.65\\
    3844&IC1613& 7496&  WFPC2&20.40&20.37&20.46&     &     &     & 1.50& 1.46& 1.53&20.37&20.33&20.43& 1.52& 1.48& 1.55&0.025&24.38&24.34&24.45\\
    3974&UGC685&10210&    ACS&24.45&24.43&24.47& 1.18& 1.17& 1.18&     &     &     &24.46&24.44&24.48& 1.53& 1.53& 1.54&0.057&28.42&28.40&28.44\\
    5818&     M33&10190&    ACS&20.89&20.84&20.98& 1.47& 1.39& 1.49&     &     &     &25.19&25.17&25.21& 1.88& 1.79& 1.91&0.051&24.82&24.76&24.92\\
    5896&NGC625& 8708&  WFPC2&24.14&24.11&24.17&     &     &     & 1.58& 1.57& 1.59&24.10&24.08&24.13& 1.60& 1.59& 1.61&0.016&28.12&28.09&28.15\\
    6430&ESO245-05& 8601&  WFPC2&24.27&24.22&24.33& 1.00& 0.98& 1.02&     &     &     &24.24&24.18&24.29& 1.43& 1.41& 1.46&0.016&28.29&28.23&28.35\\
    6699&UGC1281&10210&    ACS&24.63&24.60&24.65& 1.10& 1.10& 1.11&     &     &     &24.63&24.61&24.65& 1.44& 1.43& 1.44&0.047&28.63&28.61&28.65\\
    7671&  NGC784&10210&    ACS&24.70&24.68&24.72& 1.11& 1.10& 1.11&     &     &     &23.87&23.85&23.89& 1.45& 1.44& 1.45&0.059&28.68&28.66&28.70\\
   11139&     ESO154-23&10210&    ACS&24.77&24.75&24.79& 1.05& 1.04& 1.05&     &     &     &23.38&23.35&23.40& 1.37& 1.36& 1.37&0.015&28.84&28.82&28.86\\
   12286& NGC1313&10210&    ACS&24.36&24.34&24.37& 1.30& 1.30& 1.31&     &     &     &24.04&24.02&24.06& 1.69& 1.68& 1.69&0.111&28.21&28.19&28.22\\
   12460& NGC1311&10210&    ACS&24.69&24.67&24.71& 1.15& 1.14& 1.15&     &     &     &25.07&25.04&25.09& 1.49& 1.49& 1.50&0.020&28.73&28.71&28.75\\
   13163&  IC1959&10210&    ACS&24.96&24.93&25.00& 1.10& 1.09& 1.10&     &     &     &25.20&25.17&25.22& 1.43& 1.43& 1.44&0.011&29.03&28.99&29.06\\
   14241&  UGCA86& 9771&    ACS&25.19&25.17&25.21& 1.91& 1.90& 1.91&     &     &     &24.68&24.66&24.70& 2.39& 2.38& 2.39&0.933&27.47&27.45&27.50\\
   15439&  UGCA92& 9771&    ACS&24.89&24.86&24.93& 1.73& 1.72& 1.73&     &     &     &23.83&23.82&23.84& 2.18& 2.18& 2.19&0.789&27.46&27.43&27.50\\
   15488&NGC1560& 8192&  WFPC2&23.89&23.83&24.00& 1.25& 1.22& 1.27&     &     &     &23.86&23.80&23.96& 1.78& 1.74& 1.81&0.187&27.57&27.50&27.68\\
   16957&UGCA105& 8601&  WFPC2&24.30&23.97&24.33& 1.30& 1.29& 1.36&     &     &     &24.27&23.94&24.30& 1.84& 1.83& 1.93&0.313&27.71&27.38&27.75\\
   18731&ESO121-20& 9771&    ACS&24.87&24.82&24.92& 1.03& 1.00& 1.04&     &     &     &24.87&24.82&24.92& 1.34& 1.31& 1.36&0.041&28.90&28.85&28.95\\
   21102& NGC2366&10605&    ACS&23.53&23.51&23.55&     &     &     & 1.42& 1.40& 1.44&23.18&23.16&23.20& 1.51& 1.50& 1.53&0.036&27.57&27.55&27.60\\
   21199&ESO059-01& 9771&    ACS&24.48&24.45&24.50& 1.31& 1.31& 1.32&     &     &     &24.49&24.47&24.51& 1.70& 1.69& 1.71&0.149&28.26&28.23&28.28\\
   21302&DDO44& 8192&  WFPC2&23.43&23.36&23.51& 1.18& 1.12& 1.22&     &     &     &23.40&23.32&23.47& 1.68& 1.60& 1.74&0.041&27.36&27.27&27.45\\
   21396& NGC2403&10523&    ACS&23.50&23.46&23.55& 1.21& 1.19& 1.23&     &     &     &23.95&23.93&23.97& 1.57& 1.54& 1.60&0.040&27.50&27.45&27.55\\
   21600& UGC3974&10210&    ACS&25.61&25.58&25.64& 1.19& 1.18& 1.20&     &     &     &23.83&23.82&23.84& 1.55& 1.54& 1.56&0.033&29.62&29.59&29.65\\
   21614&    KK65&10210&    ACS&25.53&25.49&25.57& 1.12& 1.11& 1.13&     &     &     &24.97&24.93&25.00& 1.46& 1.45& 1.47&0.032&29.55&29.51&29.59\\
   23324& UGC4305&10605&    ACS&23.64&23.62&23.65&     &     &     & 1.53& 1.51& 1.54&23.93&23.91&23.95& 1.61& 1.59& 1.62&0.032&27.67&27.65&27.69\\
   23521&  M81DwA&10605&    ACS&23.61&23.55&23.65&     &     &     & 1.52& 1.48& 1.54&27.53&27.50&27.59& 1.60& 1.57& 1.62&0.021&27.66&27.60&27.71\\
   23769&DDO52& 9771&    ACS&26.07&26.03&26.11& 1.17& 1.16& 1.18&     &     &     &26.08&26.04&26.11& 1.52& 1.51& 1.54&0.037&30.08&30.04&30.12\\
   24050& UGC4459&10605&    ACS&23.81&23.78&23.83&     &     &     & 1.59& 1.58& 1.60&22.40&22.36&22.44& 1.67& 1.66& 1.67&0.037&27.82&27.80&27.85\\
   24213&UGC4483& 8192&  WFPC2&23.59&23.42&23.74& 0.91& 0.82& 0.96&     &     &     &23.56&23.38&23.71& 1.30& 1.18& 1.37&0.033&27.62&27.42&27.79\\
   26761& NGC2915& 9288&    ACS&24.51&24.48&24.54& 1.44& 1.43& 1.46&     &     &     &23.74&23.69&23.78& 1.86& 1.83& 1.87&0.275&28.04&28.01&28.07\\
   27605& UGC5139&10605&    ACS&23.99&23.97&24.01&     &     &     & 1.54& 1.51& 1.57&25.03&24.99&25.06& 1.63& 1.60& 1.65&0.049&28.00&27.97&28.02\\
   28120& NGC2976&10915&    ACS&23.90&23.88&23.93& 1.47& 1.26& 1.49&     &     &     &23.74&23.70&23.77& 1.89& 1.63& 1.92&0.081&27.77&27.74&27.84\\
   28630& NGC3031&10523&    ACS&23.84&23.80&23.89& 1.67& 1.43& 1.88&     &     &     &23.27&23.24&23.32& 2.12& 1.84& 2.35&0.079&27.68&27.59&27.77\\
   28655&     M82&10776&    ACS&24.05&24.03&24.07&     &     &     & 2.52& 2.50& 2.53&26.22&26.17&26.28& 2.51& 2.50& 2.52&0.153&27.73&27.71&27.75\\
   28731&KDG61& 8192&  WFPC2&23.77&23.60&25.18& 1.13& 1.08& 1.42&     &     &     &23.74&23.58&25.15& 1.61& 1.55& 2.01&0.073&27.66&27.42&29.08\\
   28913&DDO70& 8601&  WFPC2&21.80&21.77&21.83& 1.00& 0.98& 1.00&     &     &     &21.76&21.73&21.79& 1.43& 1.41& 1.44&0.031&25.77&25.74&25.80\\
   29128&NGC3109& 8601&  WFPC2&21.69&21.61&21.78& 1.08& 1.05& 1.10&     &     &     &21.65&21.58&21.75& 1.55& 1.50& 1.58&0.058&25.62&25.54&25.72\\
   29146& NGC3077& 9381&    ACS&24.00&23.98&24.02&     &     &     & 2.41& 2.31& 2.48&23.56&23.55&23.58& 2.42& 2.33& 2.48&0.068&27.84&27.81&27.88\\
   29194&Antlia&10210&    ACS&21.71&21.80&21.80& 1.14& 1.10& 1.18&     &     &     &21.72&21.81&21.80& 1.49& 1.44& 1.53&0.079&25.65&25.72&25.74\\
   29231&BK5N& 5898&  WFPC2&23.90&23.83&23.99&     &     &     & 1.55& 1.49& 1.58&23.87&23.79&23.95& 1.58& 1.51& 1.60&0.063&27.81&27.73&27.90\\
   29257&KDG63& 8192&  WFPC2&23.81&23.75&23.88& 1.17& 1.14& 1.19&     &     &     &23.78&23.71&23.85& 1.67& 1.62& 1.70&0.097&27.66&27.58&27.73\\
   29284& UGC5423&10605&    ACS&25.77&25.75&25.80&     &     &     & 1.69& 1.69& 1.70&22.38&22.34&22.41& 1.76& 1.76& 1.77&0.081&29.70&29.67&29.73\\
   29388&KDG64& 8192&  WFPC2&23.80&23.73&23.87& 1.23& 1.20& 1.27&     &     &     &23.77&23.70&23.83& 1.75& 1.71& 1.80&0.054&27.70&27.62&27.77\\
   29653&DDO75& 7496&  WFPC2&21.82&21.74&21.88&     &     &     & 1.40& 1.32& 1.43&21.78&21.70&21.84& 1.42& 1.33& 1.44&0.043&25.78&25.70&25.86\\
   30664&DDO78& 8192&  WFPC2&23.67&23.59&23.74& 1.17& 1.12& 1.21&     &     &     &23.63&23.55&23.71& 1.67& 1.60& 1.73&0.021&27.63&27.55&27.72\\
   30819&  IC2574&10605&    ACS&23.94&23.92&23.97&     &     &     & 1.58& 1.56& 1.59&26.00&25.84&26.13& 1.66& 1.64& 1.67&0.035&27.96&27.94&27.99\\
   30997&   DDO82&10915&    ACS&24.02&24.01&24.04& 1.29& 1.29& 1.30&     &     &     &27.54&27.51&27.59& 1.68& 1.67& 1.68&0.041&28.00&27.98&28.02\\
   31286&BK6N& 8192&  WFPC2&23.58&23.46&23.69& 1.02& 0.93& 1.11&     &     &     &23.55&23.42&23.65& 1.47& 1.34& 1.58&0.011&27.60&27.45&27.73\\
   35286&UGC6456& 6276&  WFPC2&24.34&24.31&24.38& 1.45& 1.44& 1.46&     &     &     &24.32&24.28&24.36& 2.04& 2.03& 2.05&0.037&28.24&28.20&28.28\\
   35684&UGC6541& 8601&  WFPC2&24.10&23.99&24.22& 0.99& 0.94& 1.02&     &     &     &24.06&23.95&24.18& 1.43& 1.36& 1.46&0.018&28.12&28.01&28.25\\
   35878&NGC3741& 8601&  WFPC2&23.52&23.41&23.62& 1.01& 0.96& 1.04&     &     &     &23.48&23.37&23.58& 1.45& 1.39& 1.49&0.024&27.53&27.41&27.64\\
   36014&     ESO320-14&10235&    ACS&25.02&24.98&25.05& 1.18& 1.16& 1.19&     &     &     &24.03&24.02&24.05& 1.53& 1.52& 1.55&0.144&28.83&28.79&28.87\\
   37050&DDO99& 8601&  WFPC2&23.08&23.01&23.14& 0.92& 0.87& 0.96&     &     &     &23.04&22.97&23.11& 1.32& 1.26& 1.37&0.026&27.11&27.03&27.19\\
   38148& NGC4068& 9771&    ACS&24.14&24.12&24.16& 1.07& 1.06& 1.07&     &     &     &20.90&20.85&20.98& 1.39& 1.39& 1.40&0.020&28.20&28.18&28.22\\
   38688& NGC4144& 9765&    ACS&24.40&24.36&24.44& 1.82& 1.80& 1.83&     &     &     &24.53&24.52&24.53& 2.29& 2.27& 2.30&0.015&28.32&28.27&28.36\\
   38881&NGC4163& 9771&    ACS&23.33&23.31&23.35& 1.13& 1.13& 1.14&     &     &     &23.34&23.32&23.36& 1.48& 1.48& 1.49&0.020&27.37&27.35&27.39\\
   39032&ESO321-014& 8601&  WFPC2&23.68&23.59&23.77& 1.01& 0.97& 1.04&     &     &     &23.64&23.56&23.73& 1.45& 1.39& 1.50&0.093&27.57&27.48&27.67\\
   39058& UGC7242& 9771&    ACS&24.63&24.61&24.65& 1.20& 1.19& 1.20&     &     &     &24.75&24.73&24.77& 1.56& 1.55& 1.56&0.019&28.67&28.64&28.69\\
   39145&DDO113& 8601&  WFPC2&23.38&23.31&23.44& 1.02& 0.98& 1.04&     &     &     &23.34&23.27&23.41& 1.46& 1.41& 1.49&0.020&27.36&27.29&27.44\\
   39225&NGC4214& 6569&  WFPC2&23.38&23.35&23.42&     &     &     & 1.60& 1.59& 1.61&23.35&23.31&23.39& 1.62& 1.61& 1.63&0.022&27.35&27.31&27.39\\
   39346&NGC4236& 8601&  WFPC2&24.46&24.31&24.89& 1.07& 1.03& 1.20&     &     &     &24.42&24.28&24.85& 1.54& 1.48& 1.71&0.014&28.46&28.28&28.91\\
   39422& NGC4244&10523&    ACS&24.13&24.07&24.20& 1.21& 1.16& 1.24&     &     &     &24.06&24.05&24.08& 1.58& 1.51& 1.61&0.021&28.16&28.09&28.23\\
   39573&IC3104& 8601&  WFPC2&23.53&23.49&23.57& 1.34& 1.33& 1.36&     &     &     &23.51&23.47&23.55& 1.91& 1.89& 1.92&0.416&26.80&26.76&26.84\\
   39615& UGC7356&10905&    ACS&25.37&25.35&25.39& 1.18& 1.17& 1.19&     &     &     &23.72&23.70&23.75& 1.54& 1.53& 1.55&0.022&29.40&29.38&29.43\\
   40596& NGC4395&10905&    ACS&24.40&24.39&24.42& 1.28& 1.28& 1.29&     &     &     &27.13&27.05&27.21& 1.66& 1.66& 1.67&0.017&28.42&28.41&28.44\\
   40705& NGC4407& 9363&    ACS&27.05&27.00&27.10&     &     &     & 1.76& 1.74& 1.80&24.79&24.77&24.81& 1.82& 1.80& 1.86&0.033&31.05&30.99&31.10\\
   40791&DDO126& 8601&  WFPC2&24.44&24.37&24.51& 0.95& 0.93& 0.97&     &     &     &24.41&24.34&24.48& 1.36& 1.34& 1.39&0.014&28.48&28.41&28.55\\
   40904&DDO125& 8601&  WFPC2&23.06&23.02&23.09& 1.03& 1.01& 1.05&     &     &     &23.02&22.98&23.05& 1.48& 1.45& 1.50&0.020&27.07&27.02&27.11\\
   40973&NGC4449& 8601&  WFPC2&24.18&24.03&24.39& 1.08& 1.06& 1.12&     &     &     &24.14&23.99&24.35& 1.55& 1.51& 1.60&0.020&28.17&28.01&28.39\\
   41048&UGC7605& 8601&  WFPC2&24.33&24.22&24.46& 0.90& 0.81& 0.93&     &     &     &24.29&24.18&24.42& 1.29& 1.17& 1.34&0.014&28.38&28.26&28.53\\
   42408& NGC4605& 9771&    ACS&24.74&24.72&24.76& 1.24& 1.23& 1.25&     &     &     &24.04&24.02&24.05& 1.61& 1.60& 1.62&0.014&28.78&28.75&28.80\\
   42656&IC3687& 8601&  WFPC2&24.28&24.21&24.35& 0.98& 0.96& 1.00&     &     &     &24.24&24.17&24.31& 1.41& 1.37& 1.44&0.019&28.30&28.23&28.38\\
   42936&ESO381-018& 9771&    ACS&24.67&24.62&24.72& 1.08& 1.07& 1.09&     &     &     &24.67&24.63&24.72& 1.41& 1.40& 1.42&0.063&28.65&28.59&28.69\\
   43048&     ESO381-20&10235&    ACS&24.67&24.62&24.71& 0.99& 0.99& 1.00&     &     &     &24.03&24.01&24.04& 1.30& 1.29& 1.31&0.066&28.66&28.61&28.70\\
   43495& NGC4736&10523&    ACS&24.44&24.43&24.45& 1.90& 1.90& 1.91&     &     &     &24.74&24.72&24.76& 2.39& 2.38& 2.39&0.017&28.34&28.32&28.35\\
   43869&NGC4789A&10905&    ACS&23.94&23.91&23.97& 0.98& 0.96&  1.0&     &     &     &24.36&24.34&24.37& 1.29& 1.25& 1.31&0.009&28.03&28.00&28.07\\
   43978&     ESO443-09&10235&    ACS&24.88&24.80&24.95& 1.14& 0.94& 1.20&     &     &     &24.78&24.76&24.79& 1.48& 1.23& 1.56&0.065&28.85&28.75&28.96\\
   44182&     M64&10905&    ACS&24.49&24.48&24.49& 1.93& 1.93& 1.94&     &     &     &25.29&25.27&25.31& 2.42& 2.41& 2.42&0.040&28.34&28.33&28.34\\
   44491&DDO155& 5915&  WFPC2&22.71&22.59&22.80&     &     &     & 1.47& 1.45& 1.49&22.67&22.56&22.76& 1.49& 1.47& 1.51&0.026&26.69&26.57&26.78\\
   45104&ESO269-37& 8601&  WFPC2&23.80&23.68&23.91& 1.26& 1.22& 1.34&     &     &     &23.77&23.65&23.87& 1.80& 1.75& 1.90&0.132&27.56&27.42&27.68\\
   45314&IC4182& 9162&  WFPC2&24.21&24.15&24.28& 1.07& 1.06& 1.09&     &     &     &24.17&24.11&24.24& 1.53& 1.52& 1.56&0.014&28.20&28.13&28.27\\
   45372& UGC8201&10605&    ACS&24.34&24.33&24.36&     &     &     & 1.41& 1.39& 1.42&24.89&24.81&24.96& 1.50& 1.49& 1.52&0.024&28.41&28.39&28.43\\
   45506& UGC8215& 9771&    ACS&24.24&24.19&24.29& 1.12& 1.10& 1.13&     &     &     &23.91&23.88&23.93& 1.47& 1.44& 1.48&0.011&28.30&28.25&28.36\\
   45717&ESO269-058&10235&    ACS&24.06&24.05&24.07& 1.33& 1.32& 1.33&     &     &     &24.08&24.06&24.09& 1.72& 1.71& 1.72&0.107&27.91&27.90&27.93\\
   45849& NGC5023& 9765&    ACS&24.27&24.22&24.32& 1.76& 1.74& 1.79&     &     &     &23.57&23.55&23.59& 2.23& 2.20& 2.25&0.018&28.19&28.14&28.25\\
   45916&     ESO269-66&10235&    ACS&23.96&23.94&23.97& 1.35& 1.34& 1.35&     &     &     &23.53&23.51&23.55& 1.74& 1.73& 1.75&0.097&27.82&27.80&27.84\\
   45939&DDO167& 8601&  WFPC2&24.11&23.97&24.26& 1.03& 0.98& 1.07&     &     &     &24.08&23.93&24.23& 1.47& 1.40& 1.53&0.011&28.13&27.98&28.30\\
   46039&DDO168& 8601&  WFPC2&24.28&24.22&24.34& 0.94& 0.92& 0.96&     &     &     &24.24&24.18&24.30& 1.36& 1.33& 1.38&0.014&28.32&28.25&28.39\\
   46127& UGC8331&10905&    ACS&24.13&24.09&24.16& 0.99& 0.96& 1.01&     &     &     &24.09&23.94&24.19& 1.30& 1.26& 1.32&0.009&28.22&28.17&28.26\\
   46663&KK196& 9771&    ACS&24.03&23.97&24.07& 1.17& 1.06& 1.20&     &     &     &24.03&23.98&24.07& 1.53& 1.38& 1.56&0.084&27.95&27.90&28.02\\
   46674&NGC5102& 8601&  WFPC2&23.86&23.83&23.89& 1.37& 1.36& 1.38&     &     &     &23.83&23.80&23.87& 1.94& 1.93& 1.96&0.055&27.72&27.69&27.75\\
   46680&   KK197& 9771&    ACS&24.14&24.12&24.16& 1.49& 1.47& 1.50&     &     &     &24.67&24.65&24.70& 1.91& 1.88& 1.92&0.153&27.88&27.86&27.91\\
   46885&KK200& 8192&  WFPC2&24.48&24.43&24.54& 1.15& 1.14& 1.18&     &     &     &24.45&24.39&24.50& 1.65& 1.62& 1.69&0.070&28.36&28.30&28.42\\
   46957&NGC5128& 6230&  WFPC2&24.04&24.03&24.05&     &     &     & 2.32& 2.30& 2.33&24.03&24.02&24.04& 2.35& 2.34& 2.36&0.115&27.72&27.71&27.73\\
   47073&  IC4247&10235&    ACS&24.56&24.53&24.60& 1.07& 1.06& 1.07&     &     &     &24.67&24.63&24.71& 1.40& 1.39& 1.40&0.065&28.54&28.51&28.58\\
   47171&ESO324-24& 8601&  WFPC2&24.00&23.95&24.04& 1.05& 1.03& 1.08&     &     &     &23.96&23.92&24.01& 1.51& 1.48& 1.54&0.112&27.84&27.79&27.89\\
   47495&UGC8508& 8601&  WFPC2&23.09&23.01&23.16& 0.96& 0.93& 0.98&     &     &     &23.05&22.98&23.12& 1.38& 1.34& 1.41&0.015&27.12&27.04&27.20\\
   47762&NGC5206& 6814&  WFPC2&23.38&23.36&23.40&     &     &     & 2.28& 2.26& 2.30&23.37&23.35&23.39& 2.32& 2.29& 2.34&0.120&27.06&27.03&27.08\\
   47853& NGC5238&10905&    ACS&24.20&24.18&24.22& 1.12& 1.12& 1.12&     &     &     &23.94&23.92&23.96& 1.46& 1.46& 1.47&0.009&28.27&28.25&28.29\\
   48029&     ESO444-78&10235&    ACS&24.65&24.62&24.68& 1.09& 1.08& 1.10&     &     &     &24.94&24.91&24.96& 1.43& 1.41& 1.44&0.052&28.65&28.62&28.68\\
   48082& NGC5236&10523&    ACS&24.52&24.49&24.55& 1.42& 1.39& 1.43&     &     &     &23.52&23.48&23.57& 1.82& 1.79& 1.84&0.045&28.46&28.43&28.50\\
   48139&NGC5237& 9771&    ACS&23.76&23.74&23.77& 1.32& 1.32& 1.33&     &     &     &23.77&23.76&23.79& 1.71& 1.70& 1.71&0.099&27.63&27.61&27.64\\
   48280& UGC8638& 9771&    ACS&24.11&24.08&24.13& 1.13& 1.12& 1.13&     &     &     &23.85&23.81&23.90& 1.47& 1.46& 1.48&0.013&28.16&28.14&28.19\\
   48332&UGC8651&10210&    ACS&23.37&23.33&23.41& 1.03& 1.02& 1.05&     &     &     &23.37&23.34&23.41& 1.35& 1.33& 1.37&0.006&27.46&27.42&27.50\\
   48368&IC4316& 8601&  WFPC2&24.34&24.27&24.41& 1.23& 1.20& 1.26&     &     &     &24.31&24.24&24.38& 1.75& 1.72& 1.79&0.054&28.24&28.16&28.31\\
   48467&NGC5264& 8601&  WFPC2&24.45&24.39&24.51& 1.21& 1.19& 1.23&     &     &     &24.42&24.36&24.48& 1.73& 1.70& 1.75&0.050&28.36&28.30&28.43\\
   48515&KK211& 8192&  WFPC2&23.99&23.92&24.06& 1.24& 1.22& 1.27&     &     &     &23.96&23.89&24.03& 1.77& 1.73& 1.81&0.109&27.79&27.71&27.87\\
   48738&ESO325-11& 8601&  WFPC2&23.81&23.75&23.87& 0.97& 0.95& 1.00&     &     &     &23.77&23.71&23.83& 1.40& 1.36& 1.43&0.088&27.72&27.65&27.79\\
   49158&UGC8760&10210&    ACS&23.53&23.48&23.57& 1.05& 1.01& 1.06&     &     &     &23.53&23.49&23.57& 1.38& 1.33& 1.39&0.016&27.59&27.54&27.64\\
   49448&    HOLMBERGIV&10905&    ACS&25.24&25.23&25.26& 1.19& 1.18& 1.19&     &     &     &22.39&22.36&22.44& 1.54& 1.54& 1.55&0.015&29.29&29.27&29.30\\
   49452&UGC8833& 8601&  WFPC2&23.52&23.40&23.62& 0.97& 0.88& 1.01&     &     &     &23.48&23.37&23.58& 1.40& 1.26& 1.46&0.012&27.55&27.42&27.68\\
   49615&ESO384-016&10235&    ACS&24.30&24.27&24.32& 1.20& 1.19& 1.21&     &     &     &24.31&24.28&24.33& 1.56& 1.54& 1.57&0.073&28.23&28.20&28.26\\
   50961&UGC9128&10210&    ACS&22.74&22.71&22.77& 1.02& 0.98& 1.05&     &     &     &22.75&22.71&22.78& 1.34& 1.29& 1.38&0.023&26.80&26.76&26.84\\
   51472&  DDO190&10915&    ACS&23.17&23.15&23.20& 0.99& 0.98& 1.00&     &     &     &25.99&25.83&26.12& 1.30& 1.28& 1.31&0.012&27.26&27.23&27.29\\
   51659&PGC51659& 8601&  WFPC2&23.92&23.70&24.14& 1.03& 1.00& 1.09&     &     &     &23.88&23.67&24.10& 1.48& 1.43& 1.56&0.130&27.74&27.51&27.97\\
   53639&     ESO223-09&10235&    ACS&24.43&24.39&24.47& 1.78& 1.76& 1.79&     &     &     &23.21&23.19&23.24& 2.24& 2.22& 2.26&0.255&27.93&27.89&27.97\\
   54392&     ESO274-01&10235&    ACS&23.92&23.90&23.94& 1.48& 1.48& 1.49&     &     &     &23.69&23.63&23.75& 1.90& 1.90& 1.91&0.252&27.49&27.47&27.51\\
   57888&     ESO137-18&10235&    ACS&24.51&24.47&24.56& 1.75& 1.74& 1.77&     &     &     &26.20&26.15&26.27& 2.21& 2.20& 2.23&0.243&28.04&27.99&28.09\\
   60849&IC4662& 9771&    ACS&23.07&23.05&23.09& 1.26& 1.25& 1.26&     &     &     &23.08&23.07&23.10& 1.63& 1.63& 1.63&0.070&27.00&26.98&27.02\\
   65367&DDO210& 8601&  WFPC2&20.97&20.84&21.07& 0.98& 0.86& 1.02&     &     &     &20.93&20.80&21.03& 1.41& 1.25& 1.47&0.050&24.93&24.79&25.07\\
   67908&IC5152& 8192&  WFPC2&22.56&22.51&22.62& 1.09& 1.06& 1.10&     &     &     &22.53&22.47&22.58& 1.56& 1.52& 1.58&0.025&26.55&26.49&26.61\\
   69519&Tucana& 5423&  WFPC2&20.86&20.82&20.87&     &     &     & 1.53& 1.51& 1.55&20.82&20.78&20.84& 1.55& 1.53& 1.57&0.031&24.81&24.77&24.84\\
   71431&UGCA438& 8192&  WFPC2&22.72&22.64&22.78& 1.07& 1.04& 1.10&     &     &     &22.68&22.60&22.75& 1.53& 1.50& 1.57&0.015&26.72&26.63&26.79\\
   71538&Pegasus& 5915&  WFPC2&21.03&20.98&21.12&     &     &     & 1.73& 1.70& 1.77&20.99&20.95&21.08& 1.76& 1.72& 1.79&0.067&24.90&24.85&24.99\\
   73049& NGC7793&10523&    ACS&23.76&23.70&23.82& 1.21& 1.14& 1.25&     &     &     &23.96&23.94&23.98& 1.57& 1.49& 1.62&0.018&27.79&27.72&27.87\\
   86668&D564-08& 9771&    ACS&25.69&25.63&25.75& 1.11& 1.06& 1.16&     &     &     &25.70&25.64&25.76& 1.45& 1.39& 1.51&0.029&29.72&29.65&29.79\\
   86669&D565-06& 9771&    ACS&25.80&25.70&25.90& 1.05& 0.99& 1.08&     &     &     &25.80&25.71&25.90& 1.37& 1.30& 1.41&0.038&29.83&29.73&29.95\\
   95597&   KKH37& 9771&    ACS&23.72&23.67&23.75& 1.18& 1.17& 1.19&     &     &     &22.61&22.58&22.66& 1.54& 1.53& 1.55&0.074&27.65&27.61&27.69\\
  166082&CamA& 8192&  WFPC2&24.32&24.20&24.43& 1.19& 1.11& 1.21&     &     &     &24.28&24.16&24.40& 1.70& 1.58& 1.73&0.216&27.94&27.81&28.06\\
  166084&CamB& 8192&  WFPC2&24.00&23.84&24.14& 1.19& 1.14& 1.24&     &     &     &23.97&23.81&24.11& 1.70& 1.63& 1.77&0.217&27.63&27.46&27.79\\
  166101&KK77& 8192&  WFPC2&23.88&23.80&23.96& 1.29& 1.21& 1.33&     &     &     &23.85&23.78&23.93& 1.83& 1.72& 1.88&0.138&27.63&27.54&27.73\\
  166152&   KK182& 9771&    ACS&25.10&24.98&25.22& 1.05& 1.03& 1.12&     &     &     &24.89&24.80&24.96& 1.38& 1.35& 1.46&0.103&29.02&28.88&29.14\\
  166158&KK189& 9771&    ACS&24.24&24.15&24.31& 1.27& 1.22& 1.31&     &     &     &24.25&24.16&24.32& 1.65& 1.59& 1.70&0.110&28.09&28.00&28.18\\
  166175&KK217& 8192&  WFPC2&23.90&23.74&24.07& 1.29& 1.07& 1.38&     &     &     &23.87&23.71&24.03& 1.84& 1.54& 1.96&0.120&27.67&27.49&27.90\\
  166179&KK221& 8601&  WFPC2&24.28&24.22&24.34& 1.26& 1.24& 1.29&     &     &     &24.25&24.19&24.31& 1.80& 1.77& 1.83&0.139&28.02&27.96&28.09\\
  166185&KK230& 9771&    ACS&22.63&22.51&22.73& 1.01& 0.98& 1.03&     &     &     &22.63&22.51&22.73& 1.32& 1.29& 1.35&0.014&26.70&26.58&26.81\\
  490287&     ESO215-09&10235&    ACS&24.93&24.91&24.96& 1.28& 1.27& 1.28&     &     &     &23.22&23.19&23.25& 1.65& 1.65& 1.66&0.218&28.60&28.57&28.62\\
 2801026&KKR25& 8601&  WFPC2&22.35&22.28&22.41& 0.94& 0.91& 0.97&     &     &     &22.31&22.25&22.37& 1.35& 1.31& 1.40&0.009&26.40&26.33&26.47\\
 2806961&D634-03& 9771&    ACS&25.87&25.77&25.97& 1.11& 1.07& 1.13&     &     &     &25.88&25.77&25.98& 1.45& 1.39& 1.48&0.039&29.89&29.78&30.00\\
 2807103&KKH6& 9771&    ACS&24.47&24.41&24.52& 1.38& 1.37& 1.41&     &     &     &24.49&24.43&24.54& 1.78& 1.77& 1.82&0.353&27.88&27.82&27.94\\
 2807150&KKH86& 8601&  WFPC2&23.05&22.73&23.21& 0.97& 0.88& 1.01&     &     &     &23.01&22.69&23.17& 1.39& 1.27& 1.46&0.027&27.07&26.73&27.25\\
 2807155&CasdSph& 8192&  WFPC2&20.86&20.78&20.92& 1.30& 1.24& 1.36&     &     &     &20.83&20.75&20.89& 1.85& 1.76& 1.94&0.199&24.50&24.40&24.58\\
 2815822&KKs55&10235&    ACS&24.08&24.04&24.11& 1.30& 1.26& 1.33&     &     &     &24.09&24.05&24.12& 1.69& 1.64& 1.72&0.144&27.87&27.83&27.91\\
 2815823&KKs57& 9771&    ACS&24.00&23.83&24.26& 1.36& 1.16& 1.52&     &     &     &24.02&23.85&24.26& 1.75& 1.51& 1.94&0.091&27.88&27.67&28.17\\
 3097691&Cetus& 8601&  WFPC2&20.99&20.73&22.17& 0.97& 0.90& 1.10&     &     &     &20.96&20.70&22.13& 1.40& 1.30& 1.58&0.106&24.87&24.58&26.07\\
 3097828&FM1& 8601&  WFPC2&23.88&23.81&23.94& 1.13& 1.08& 1.18&     &     &     &23.84&23.78&23.91& 1.62& 1.54& 1.68&0.081&27.75&27.67&27.83\\
99990001&F8D1& 5898&  WFPC2&24.12&24.06&24.17&     &     &     & 1.86& 1.84& 1.88&24.09&24.03&24.14& 1.88& 1.87& 1.90&0.110&27.89&27.83&27.95\\
99990002&IKN& 9771&    ACS&23.93&23.91&23.95& 1.28& 1.27& 1.30&     &     &     &23.94&23.92&23.97& 1.67& 1.65& 1.68&0.057&27.88&27.85&27.90\\
99990004&HS117& 9771&    ACS&24.07&24.03&24.11& 1.21& 1.20& 1.23&     &     &     &24.08&24.05&24.12& 1.58& 1.56& 1.60&0.113&27.94&27.89&27.97\\
99990024&CenN& 9771&    ACS&23.98&23.89&24.06& 1.45& 1.17& 1.65&     &     &     &24.00&23.92&24.10& 1.87& 2.12& 2.10&0.140&27.75&27.62&27.89\\
\enddata

\label{db}

\end{deluxetable}
\clearpage
        
        \end{landscape}

\end{document}